\documentclass{article}
\usepackage[top=1in, bottom=1in, left=1in, right=1in]{geometry}
\usepackage{graphicx} 
\usepackage{subcaption} 
\usepackage{multirow}
\usepackage{array}
\usepackage{amsmath} 
\usepackage{amssymb}
\usepackage{setspace}
\usepackage{colortbl}
\usepackage[table]{xcolor} 
\usepackage{pifont}
\usepackage{verbatim}
\usepackage{float}
\usepackage{tabularx} 
\usepackage{placeins} 
\usepackage{tcolorbox}
\usepackage{url}

\title{Evaluating ChatGPT-3.5 Efficiency in Solving Coding Problems of Different Complexity Levels: An Empirical Analysis}
\author{Minda Li and Bhaskar Krishnamachari \\ \\ \{mcli, bkrishna\}@usc.edu \\ Department of Electrical and Computer Engineering \\ University of Southern California, Los Angeles, CA, USA}

\begin{document}

\maketitle

\begin{abstract}

ChatGPT and other large language models (LLMs) promise to revolutionize software development by automatically generating code from program specifications. We assess the performance of ChatGPT’s GPT-3.5-turbo model on LeetCode, a popular platform with algorithmic coding challenges for technical interview practice, across three difficulty levels: easy, medium, and hard. We test three main hypotheses: First, that ChatGPT solves fewer problems as difficulty rises (Hypothesis 1). Second, that prompt engineering improves ChatGPT’s performance, with greater gains on easier problems and diminishing returns on harder ones (Hypothesis 2). Third, that ChatGPT performs better in popular languages like Python, Java, and C\texttt{++} than in less common ones like Elixir, Erlang, and Racket (Hypothesis 3). To investigate these hypotheses, we conduct automated experiments using Python scripts to generate prompts that instruct ChatGPT to create Python solutions. These solutions are stored and manually submitted on LeetCode to check their correctness. For Hypothesis 1, results show the GPT-3.5-turbo model successfully solves 92\% of easy, 79\% of medium, and 51\% of hard problems. For Hypothesis 2, prompt engineering yields improvements: 14-29\% for Chain of Thought Prompting, 38-60\% by providing failed test cases in a second feedback prompt, and 33-58\% by switching to GPT-4. From a random subset of problems ChatGPT solved in Python, it also solved 70\% in Java, 50\% in C\texttt{++}, and none in Elixir, Erlang, or Racket. These findings generally validate all three hypotheses.
\end{abstract}


\section{Introduction}


Large language models have catalyzed significant advancements in various domains, including natural language processing \cite{zhou2023large}, text generation \cite{li2022pretrained}, and now code generation \cite{li2023think}. LLMs have demonstrated skills in understanding, generating, and manipulating code, transforming the landscape of programming assistance. Large language models are trained on vast datasets comprising both natural language text and source code, which enables them to offer support in software development tasks. One of the more prominent LLMs is ChatGPT, a conversational AI model developed by OpenAI. The potential of these models to assist with programming has led to increased interest in their ability to improve programmer productivity, code correctness, and automated code generation from problem specifications. 
 

In this paper, we focus on exploring ChatGPT's ability (primarily working with the low-cost GPT-3.5-turbo model)  to solve coding problems in the popular programming language Python, using the platform Leetcode~\cite{leetcode}. Leetcode is an online platform with more than 2000 coding challenges widely used for technical interview preparation and programming practice. The platform categorizes problems into three difficulty levels — easy, medium, and hard — covering a broad range of topics such as algorithms, databases, shell scripting, and concurrency. Each problem provides a detailed problem statement and typically includes 3-4 example input-output pairs. LeetCode’s integrated compiler and test cases make it an ideal tool for evaluating the correctness of solutions generated by GPT models, allowing for a structured and objective comparison. It is important to note that Leetcode, while useful as a benchmarking tool, does not fully capture the complexities of real-world coding environments. 


Our research is guided by three primary questions aimed at assessing the strengths and limitations of ChatGPT in coding tasks:

\textbf{1. How effectively can ChatGPT solve Leetcode problems across varying levels of difficulty (easy, medium, and hard)?}

\textbf{2. Can prompt engineering and the selection of more effective demonstrations enhance ChatGPT's initial coding results?}

\textbf{3. Which programming languages is ChatGPT most proficient in?}


One key challenge with language models like ChatGPT is the phenomenon of ``hallucinations" - where the model may guess the user's intent rather than seek clarification, respond sensitively to input phrasing, or produce plausible yet incorrect code \cite{maynez2020faithfulness}. These issues may arise from the limitations in the data set on which GPT-3.5-turbo was trained, particularly with respect to the code. LeetCode provides a robust framework of benchmarks and test cases that a solution must pass to be deemed correct. Even minor misunderstandings by ChatGPT can cause the generated code to fail in fulfilling its intended purpose. Given ChatGPT's sensitivity to prompt phrasing, we also explore how the model's performance can be improved through ``prompt engineering": designing more precise and effective prompts. In addition to Python, we examine ChatGPT’s proficiency across various programming languages, from popular ones like C++ and Java to less common languages such as Erlang, Elixir, and Racket. Through these investigations, we seek to understand better the strengths and limitations of ChatGPT’s code-generation capabilities.


To conduct our evaluation, we developed a python script that interacts with ChatGPT via an API key, storing the model's responses in corresponding files for each query. For each coding problem, we instructed ChatGPT to provide its solutions in python to ensure consistency in evaluation across all test cases. We tested 1,475 LeetCode coding problems, for each entering ChatGPT responses into the platform, and recording the result, including the number of passed test cases. These results were then analyzed to evaluate ChatGPT's problem-solving performance and further test specific hypotheses.


With respect to Question 1, our results show that ChatGPT 3.5 turbo performed well in solving LeetCode problems, with a high success rate of 92\% for easy problems and 79\% for medium ones. However, when tested on LeetCode's hard category, the model's performance dropped, achieving a 51\% pass rate. These results highlight the model's competence in simpler tasks, but also reveal its challenges when handling more complex coding problems.

Regarding Question 2, our evaluation revealed that while chain-of-thought (CoT) prompt engineering improved ChatGPT performance in all problem difficulties, its impact was most significant on easier problems, with less impactful results on medium and hard problems. In contrast, incorporating test cases into the query — where the GPT-3.5-turbo model initially produced incorrect results — along with the expected outputs, provided even more significant improvements, particularly for medium and hard problems. Lastly, we observed that ChatGPT using the more advanced GPT-4 model, when paired with CoT prompting, also delivered strong performance, closely matching the results of adding in test cases where GPT-3.5-turbo initially failed, especially on more complex problems. This indicates that CoT prompting combined with error-focused adjustments can be highly effective, especially as the complexity of the problem increases. For example, in the easy category of problems, we found that CoT gave 29\% improvements over the baseline, while providing failed test-cases gave 38\% improvement, and switching to GPT-4 gave 33\% improvement. For the medium and hard categories the performance of using GPT-4 and GPT-3.5-turbo with failed cases were more comparable. 

Lastly, looking at Question 3, we evaluated the performance of ChatGPT in five additional programming languages besides Python: C++, Java, Erlang, Elixir, and Racket. Python served as the baseline for comparison throughout our exploration of this question. ChatGPT was able to solve in C++ about 50\% of a random sample of problems it could solve using Python and in Java it could solve about 70\% of a random sample of problems it could solve using Python. Interestingly, some problems that ChatGPT failed to solve in Python were successfully solved in C++ or Java, indicating slight performance variations between these languages. However, for less common languages such as Erlang, Elixir, and Racket, the performance of ChatGPT was poor, with no problems successfully solved, regardless of whether they were solved in Python or not. This may be attributed to the lower frequency of these languages in the model's training data set. 

We also present results from some further experiments where we evaluated which types of problems ChatGPT performed best/worst at, and whether there is a correlation between lines of code produced by ChatGPT and the correctness of the program. 


The subsequent sections of this paper dive deeper into the context, methodology, and outcomes of our study, providing a comprehensive analysis of ChatGPT's performance and its implications for code generation tasks. In section~\ref{sec:rw}, we provide a comprehensive review of the relevant literature on large language models (LLMs) and their application to code generation, highlighting prior research and existing challenges in the field. In section~\ref{sec:methodology}, we outline the experimental setup, including the process of data collection, the tools and techniques used to evaluate ChatGPT's performance, and the criteria for assessing the correctness of its solutions. In section~\ref{sec:results}, we present a detailed analysis of ChatGPT’s performance across different categories of coding problems and programming languages. This section includes a quantitative assessment of its accuracy, efficiency, and limitations, supported by specific examples and findings from the experiments. We present our concluding comments and ideas regarding future work directions in section~\ref{sec:conclusions}.

\section{Related Works \label{sec:rw}}

We briefly identify below some of the research literature most closely related to our study.

\textbf{LLMs used to code: } Before large language models, developers often turned to platforms like Stack Overflow and GitHub for assistance with coding issues. However with the rise of LLMs, the use of large language models to solve coding issues has become increasingly popular. LLMs have demonstrated the ability of code understanding~\cite{nam2024using}, code generation~\cite{liu2023code}, test automation~\cite{10366647}, and program repair~\cite{jin2023inferfix}. In our work, we focus on evaluating the code generation capabilities of LLMs, with a focus on GPT-3.5.turbo but also include some comparative experiments using the GPT-4, Claude 3 Sonnet, and Gemini 1.0 Pro models. 

\textbf{Prompt engineering:} As the use of LLMs has increased, the need to improve the responses of LLMs has been investigated through prompt engineering. Prompt engineering is the practice of designing inputs to produce optimal results from an LLM. A key factor in this process is the sensitivity of LLMs to the exact phrasing of prompts~\cite{wei2023chainofthought}. One particular approach, chain-of-thought prompting, enables LLMs to decompose multi-step problems into a sequence of intermediate steps, improving their performance on complex tasks~\cite{yu2023towards}. In this work, we attempt to optimize ChatGPT's ability to solve coding problems through chain-of-thought prompting \cite{wei2023chainofthought} and selecting better demonstration examples~\cite{chen2021evaluating}. 

\textbf{Struggles of LLMs:} Despite their advancements, large language models (LLMs) still face significant challenges. First, the scope of knowledge available to an LLM is limited by its training data. For instance, ChatGPT's GPT-4 model, released in March 2023, builds on prior iterations like GPT-3.5-turbo, and has knowledge up until January 2022. Consequently, LLMs are susceptible to reflecting biases from the datasets they were trained on \cite{shaikh2023second}. Additionally, ChatGPT can struggle to correct errors based on provided feedback, particularly in coding contexts, due to its training limitations \cite{sakib2023extending}. Another challenge is inefficiency in algorithmic code generation \cite{Pan2023MeasuringEC}. One of ChatGPT’s most significant shortcomings is hallucination, where the model generates unverifiable or incorrect statements \cite{Bang2023AMM}. This issue can lead to faulty code output, especially when ChatGPT lacks a complete understanding of the task or access to external resources. Furthermore, when presented with ambiguous or complex queries, ChatGPT often makes assumptions rather than seeking clarification, which can further exacerbate its limitations. Given these challenges, our research seeks to quantify ChatGPT’s effectiveness in solving coding problems, especially in light of its limitations in understanding and generating code. We designed our evaluation process with these issues in mind, particularly focusing on how the model handles feedback, algorithmic complexity, and ambiguity in coding tasks. By recognizing these potential pitfalls, we aim to assess both the model’s strengths and its areas for improvement, providing insights into how it might be optimized for better performance in real-world coding applications.

Although several studies have examined LLMs' performance in solving LeetCode problems \cite{bubeck2023sparks}, \cite{nascimento2023comparingsoftwaredeveloperschatgpt}, and \cite{Nikolaidis:2023}, they have all explored less than 250 problems, unlike our more comprehensive dataset of 1475 problems; moreover we provide results of evaluations on more languages as well as some comparisons with other models. Other coding platforms, such as HumanEval \cite{Adamson2023AssessingTE}, \cite{chen2021evaluating} and CodeContests \cite{li2023explaining}, have also been tested with LLMs, highlighting their potential across a wider range of challenges. The table below compares the scope and methodology of three closely related studies with our own work. The left column outlines the areas our study investigates, while a checkmark indicates that a study fully addresses a specific topic, and an ``X" signifies that it does not.

\begin{table}[H]
    \centering
    \resizebox{\textwidth}{!}{%
    \begin{tabular}{|>{\centering\arraybackslash}p{0.3\linewidth}|>{\centering\arraybackslash}p{0.3\linewidth}|>{\centering\arraybackslash}p{0.3\linewidth}|>{\centering\arraybackslash}p{0.3\linewidth}|}
        \hline
        \large \textbf{Our Paper} & \large \textbf{Bucaioni et al. \cite{BUCAIONI2024100526}} & \large \textbf{Tian et al. \cite{tian2023chatgpt}} & \large \textbf{Hans \cite{unknown}} \\
        \hline
        \textbf{More than 1000 problems} & X: 240 problems & X: 240 problems & X: 140 problems \\
        \hline
        \textbf{3 levels} & \checkmark & \checkmark & X: Easy and Medium \\
        \hline
        \textbf{Use of multiple LLMs} & X & \checkmark & \checkmark \\
        \hline
        \textbf{Python} & X & \checkmark & \checkmark\\
        \hline
        \textbf{Other Languages} & \checkmark: C++ and Java & X & X \\
        \hline
        \textbf{Prompt Engineering} & X & X & \checkmark\\
        \hline
    \end{tabular}%
    }
    \caption{Comparison of 3 closely related papers evaluating GPT model performance on LeetCode}
    \label{tab:secondtable}
\end{table}


\section{Methodology \label{sec:methodology}}

We present here the methodology for our empirical analysis. We have made the corresponding source code publicly available at \url{https://github.com/ANRGUSC/coding_gpt_testing}

\subsection{General Methodology: }
To begin, we first needed to determine which large language model (LLM) to use for testing coding problems. We compared the performance of ChatGPT's GPT-3.5-turbo and Google’s Bard by evaluating both models on the first ten easy and medium Leetcode problems. GPT-3.5-turbo significantly outperformed Bard, leading us to focus our study primarily on the GPT-3.5-turbo model. 

A total of 1,475 Leetcode problems were selected for testing. These problems, available on LeetCode, were taken up to problem number 1825, excluding all premium-level questions. To speed up the testing process, we automated the creation of the 1,475 queries that would be sent to ChatGPT. We used a LeetCode CSV file to streamline the query writing process. All queries were named query\{problem number\}.txt. The queries followed a consistent structure, beginning with: ``Write a Python program to do the following, and place this program in a JSON object as the value corresponding to the key `code'." The problem description was then inserted, including an explanation of the problem and 2-3 test cases that the program should be able to solve. This was followed by the instruction: ``Write code that fits the following template: The program should continue below the following lines."

Lastly, the two initial lines of code provided by LeetCode needed to be inserted into the end of each query. To streamline this process, a curl program was written to download the source code for all 1,475 problems. Then another program was created to execute the curl command for each problem listed in the Leetcode CSV file. A parser program was also implemented to read each curl file, and another script ran the parser program to extract the two lines of code from each problem. These two lines were subsequently added to the queries.  Figure~\ref{fig:query1} shows an example of a written query. 

\begin{figure}[htb]
  \centering
  \begin{tcolorbox}[colback=gray!10, colframe=black!75, width=\linewidth, boxrule=0.5pt, sharp corners, fontupper=\footnotesize]
\begin{verbatim}
Write a python program to do the following, and put this program in a json object as the value 
corresponding to the key `code'. Given an array of integers `nums` and an integer `target', return indices 
of the two numbers such that they add up to `target'.

You may assume that each input would have exactly one solution, and you may not use the same element twice.

You can return the answer in any order.


Example 1:
Input: nums = [2,7,11,15], target = 9
Output: [0,1]
Output: Because nums[0] + nums[1] == 9, we return [0, 1].

Example 2:
Input: nums = [3,2,4], target = 6
Output: [1,2]

Example 3:
Input: nums = [3,3], target = 6
Output: [0,1]

Constraints:
`2 <= nums.length <= 103'
`-109 <= nums[i] <= 109'
`-109 <= target <= 109'
Only one valid answer exists.
Write code that fits the following template: 
The program should continue below the following lines:
class Solution:
    def twoSum(self, nums: List[int], target: int) -> List[int]:
        
\end{verbatim}
  \end{tcolorbox}
  \caption{Initial Query Example} 
  \label{fig:query1}
\end{figure}

Once all queries were written, we developed a Python script to interact with the OpenAI GPT-3.5-turbo API. This script submitted each query (at a temperature of 0.7) and saved ChatGPT’s response to a file named code\_query\{problem number\}.py.  

Next, we tested all generated code on Leetcode. For a solution to be accepted, it had to pass all of Leetcode’s test cases, which ranged from 10 to 100 per problem. If a solution was not accepted, the number of test cases passed served as a metric to assess how close ChatGPT was to correctly solving the problem.


\subsection{Main Hypotheses}
As we continued to test the problems from Leetcode, new questions and hypotheses emerged in regard to each of our research questions, guiding further exploration and testing. Each hypothesis is in respect to the three research questions we have stated earlier. 

\textbf{Hypothesis 1:} As the difficulty of problems increases, ChatGPT will solve fewer test cases.

\textbf{Hypothesis 2:} ChatGPT's performance will improve with the use of prompt engineering techniques. We hypothesized that prompt engineering would yield the most significant improvements for easy problems, with diminishing returns for harder problems.

\textbf{Hypothesis 3:} ChatGPT's performance will vary based on the programming language used. We hypothesized that ChatGPT would perform similarly in well-established languages like Java and C++ compared to Python, but results would differ for less common languages like Elixir, Erlang, and Racket. 

\subsection{Hypothesis 1: }
To test Hypothesis 1, we examined if a problem passed all the test cases in LeetCode. We noted if a problem was correctly solved or not and then grouped the results based on the difficulty level — easy, medium, or hard. First, problems were pre-categorized into these three difficulty levels based on their complexity and structure. Then, GPT-3.5-turbo was prompted with each problem in the dataset, and the number of test cases it solved successfully was recorded. After collecting the results, we grouped the number of correct solutions by difficulty level to enable comparison. We measured performance by calculating the percentage of test cases solved within each category, which allowed us to evaluate how the model's effectiveness varied with problem difficulty. By comparing ChatGPT's performance across these levels, we aimed to determine whether the hypothesis holds true — that the model solves fewer test cases as the difficulty of the problems increases.

\subsection{Hypothesis 2: }
Next, for testing Hypothesis 2, we sought to evaluate if it is possible to improve ChatGPT’s performance by refining the prompts. We evaluated the following three possible improvements: 

\textbf{Improvement 1: }The first improvement method applied was chain-of-thought (CoT) prompt engineering. We first began to go back and prompt engineer the initial queries written. The first method completed was chain of thought prompt engineering. We selected a subset of 20 unsolved problems and 10 solved problems from each difficulty level (easy, medium, hard). The 20 unsolved problems were chosen based on the lowest acceptance rates, while the solved problems were randomly selected. As difficulty increased, a greater number of problems had an acceptance rate of 0\%, which means that no test cases were passed. 
In order to execute CoT prompt enginering on each query, we asked ChatGPT to first generate pseudocode before writing the actual program. The pseudocode step involved analyzing the input, desired output, and test cases. If any issues were identified, ChatGPT was asked to correct the pseudocode before proceeding to the Python code. Here’s an example of the modified prompt:
``You are asked to write a Python program in a two-step process. First, write pseudocode for this program in natural language, taking into account the input, desired output, and the examples provided. Analyze your pseudocode for any mistakes and fix them. In the second step, write the Python program and put it in a JSON object under the key `code'." 
Figure~\ref{fig:query640} shows an example of a written query. 
\begin{figure}[H]
  \centering
  \begin{tcolorbox}[colback=gray!10, colframe=black!75, width=\linewidth, boxrule=0.5pt, sharp corners, fontupper=\footnotesize]
\begin{verbatim}
You are asked to write a python program to do the following, as a two-step process. First write a 
pseudo-code for this program in natural language, taking into account what is the input, what is the 
desired output and look at the examples given. Analyze your pseudo-code for any possible mistakes and fix 
them in the following second step. In the second step, write a python program to do the following, and put 
this program in a json object as the value corresponding to the key `code'.

Solve a given equation and return the value of `x' in the form of string "x=#value". The equation contains 
only `+', `-' operation, the variable `x' and its coefficient.

If there is no solution for the equation, return "No solution".

If there are infinite solutions for the equation, return "Infinite solutions".

If there is exactly one solution for the equation, we ensure that the value of `x' is an integer.


Example 1:
Input: "x+5-3+x=6+x-2"
Output: "x=2"

Example 2:
Input: "x=x"
Output: "Infinite solutions"

Example 3:
Input: "2x=x"
Output: "x=0"

Example 4:
Input: "2x+3x-6x=x+2"
Output: "x=-1"

Example 5:
Input: "x=x+2"
Output: "No solution"

Write code that fits the following template. Please put this program in a json object as the value 
corresponding to the key `code'. 

The program should continue below the following lines:
class Solution:
    def solveEquation(self, equation: str) -> str:
        
\end{verbatim}
  \end{tcolorbox}
  \caption{Chain of Thought Prompt Engineering Query Example} 
  \label{fig:query640}
\end{figure}

\textbf{Improvement 2:}
When testing on Leetcode, the system indicated which test cases were not passed. In the second improvement method, we incorporated these failed test cases into the prompts. The same set of problems used in the Chain of Thought Prompt Engineering was tested. Our hypothesis was that explicitly providing the failed test cases might help ChatGPT refine its solutions. For each query, we included the failed test case as follows:
``Here is a test case the previous program failed to solve. Input: grid = [[0,0,0,0,0,1],[1,1,0,0,1,0],[0,0,0,0,1,1],[0,0,1,0,1,0],[0,1,1,0,0,0],[0,1,1,0,0,0]]. Output: -1. Expected: 11. Please consider this when writing the next program."
Figure~\ref{fig:query640error} shows an example of a written query. 

\begin{figure}[H]
  \centering
  \begin{tcolorbox}[colback=gray!10, colframe=black!75, width=\linewidth, boxrule=0.5pt, sharp corners, fontupper=\footnotesize]
\begin{verbatim}
Write a python program to do the following please, and put this program in a json object as the value 
corresponding to the key `code'.

Solve a given equation and return the value of `x' in the form of string "x=#value". The equation contains 
only `+', `-' operation, the variable `x' and its coefficient.

If there is no solution for the equation, return "No solution".

If there are infinite solutions for the equation, return "Infinite solutions".

If there is exactly one solution for the equation, we ensure that the value of `x' is an integer.


Example 1:
Input: "x+5-3+x=6+x-2"
Output: "x=2"

Example 2:
Input: "x=x"
Output: "Infinite solutions"

Example 3:
Input: "2x=x"
Output: "x=0"

Example 4:
Input: "2x+3x-6x=x+2"
Output: "x=-1"

Example 5:
Input: "x=x+2"
Output: "No solution"

Here is a test case that the previous program failed because ValueError: invalid literal for int() with 
base 10: `+' constant += int(term) Line 58 in get_coefficient_and_constant (Solution.py) left_coefficient, 
left_constant = self.get_coefficient_and_constant(left_side) Line 9 in solveEquation (Solution.py) 
ret = Solution().solveEquation(param_1) Line 79 in _driver (Solution.py) _driver() Line 90 in <module> 
(Solution.py). Please write code with this test case in mind, and error in mind.  

Write code that fits the following template. Please put this program in a json object as the value 
corresponding to the key `code'. 

The program should continue below the following lines:
class Solution:
    def solveEquation(self, equation: str) -> str:
        
\end{verbatim}
  \end{tcolorbox}
  \caption{Inputting Failed Test Cases Query Example} 
  \label{fig:query640error}
\end{figure}

\textbf{Improvement 3:}
During our study, a more advanced version of ChatGPT, namely, GPT-4 was released. To evaluate its performance relative to GPT-3.5-turbo, we modified our testing code to call GPT-4 instead. The same set of problems used in Chain of Thought Prompt Engineering was tested. This allowed us to compare the two versions of ChatGPT on identical problem sets. 



\subsection{Hypothesis 3: }
Lastly, to test Hypothesis 3, we further evaluated Leetcode problems in five additional languages: Java, C++, Elixir, Erlang, and Racket. Specifically, we selected 10 problems that were successfully solved in Python and 10 that were not. This allowed us to use Python as a baseline for comparison with the other five languages. The decision to include these particular languages was motivated by their diversity in terms of syntax, paradigms, and popularity, allowing us to assess the adaptability of ChatGPT to both mainstream and less common programming languages.

The only modification made when testing this hypothesis was specifying which language ChatGPT-3.5-turbo should write in when generating its response in the query. We ensured that the problem descriptions remained consistent across all tests to avoid any variation in how the model interpreted the tasks. Figure~\ref{fig:query3335} shows an example of a written query. 
\begin{figure}[H]
  \centering
  \begin{tcolorbox}[colback=gray!10, colframe=black!75, width=\linewidth, boxrule=0.5pt, sharp corners, fontupper=\footnotesize]
\begin{verbatim}
Write a C++ program to do the following please, and put this program in a json object as the value 
corresponding to the key `code'.

You are given an array of integers `distance'.

You start at point `(0,0)' on an X-Y plane and you move `distance[0]' meters to the north, then 
`distance[1]' meters to the west, `distance[2]' meters to the south, `distance[3]' meters to the east, and 
so on. In other words, after each move, your direction changes counter-clockwise.

Return `true' if your path crosses itself, and `false' if it does not.


Example 1:
Input: distance = [2,1,1,2]
Output: true

Example 2:
Input: distance = [1,2,3,4]
Output: false

Example 3:
Input: distance = [1,1,1,1]
Output: true

Constraints:
`1 <= distance.length <= 500'
`1 <= distance[i] <= 500'
Follow up: Could you write a one-pass algorithm with `O(1)' extra space?

Write code that fits the following template. The program should continue between the following lines:

class Solution {
public:
    bool isSelfCrossing(vector<int>& distance) {
        
    }
};
        
\end{verbatim}
  \end{tcolorbox}
  \caption{Other Languages Query Example (C++)} 
  \label{fig:query3335}
\end{figure}

With these methodologies in place, we now turn to the results, analyzing ChatGPT's performance across varying problem difficulty levels, prompt engineering techniques, and different programming languages.

\section{Results \label{sec:results}}

In this section, we present the outcomes of our experiments, aligning them with the hypotheses outlined in the methodology. The results are organized according to the complexity of the problem, code characteristics, prompt engineering methods, programming languages, and problem types, each validated or challenged by the findings.

\subsection{Hypothesis 1 -- Difficulty Levels:}
\begin{figure}[H]
    \centering
    \includegraphics[width=\linewidth]{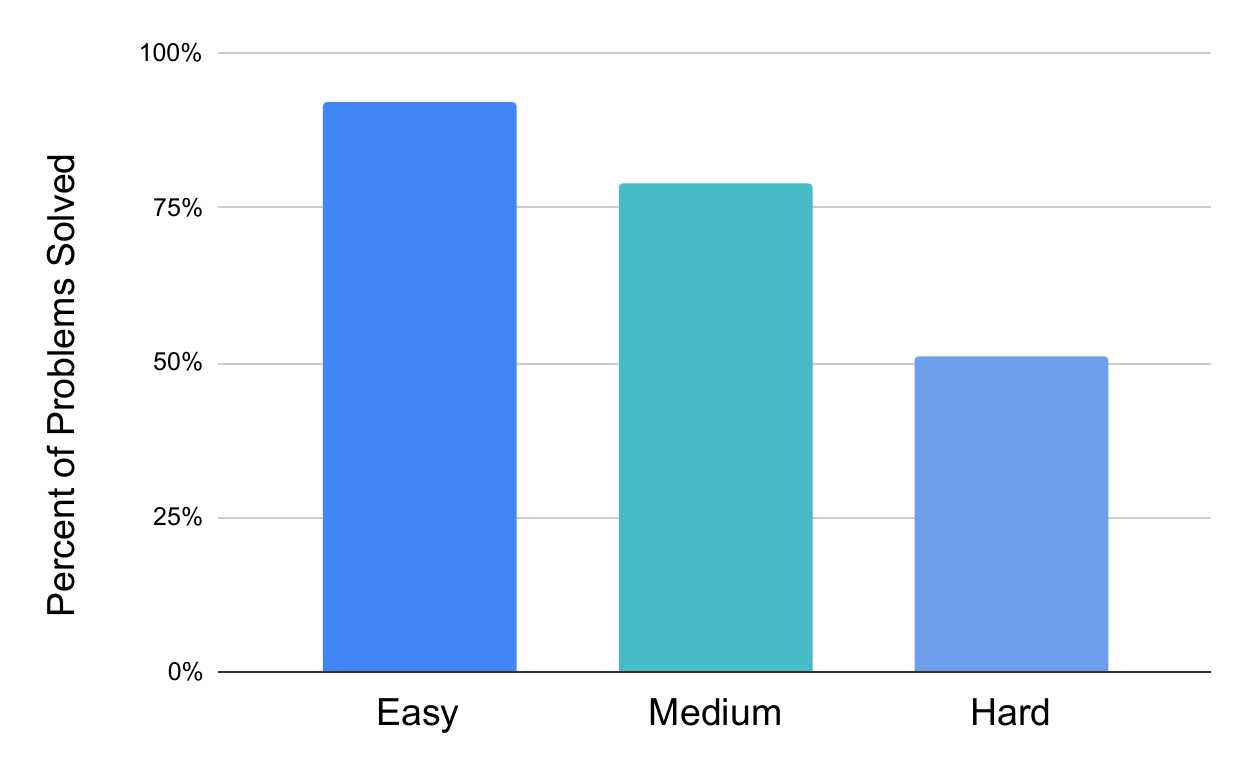}
    \caption{Results of GPT-3.5-turbo Solving Easy, Medium, and Hard Problems of the 1,475 subset}
    \label{fig:chart}
\end{figure}

We tested our first hypothesis that as the difficulty of the problem increases, ChatGPT (using GPT-3.5-turbo) would solve fewer test cases. The results supported this hypothesis: GPT-3.5-turbo successfully solved 92\% of easy problems, 79\% of medium problems, and 51\% of hard problems.
Figure~\ref{fig:chart} shows these findings.

\begin{figure}[H] 
    \centering
    \begin{subfigure}{0.32\linewidth}
        \centering
        \includegraphics[width=\linewidth]{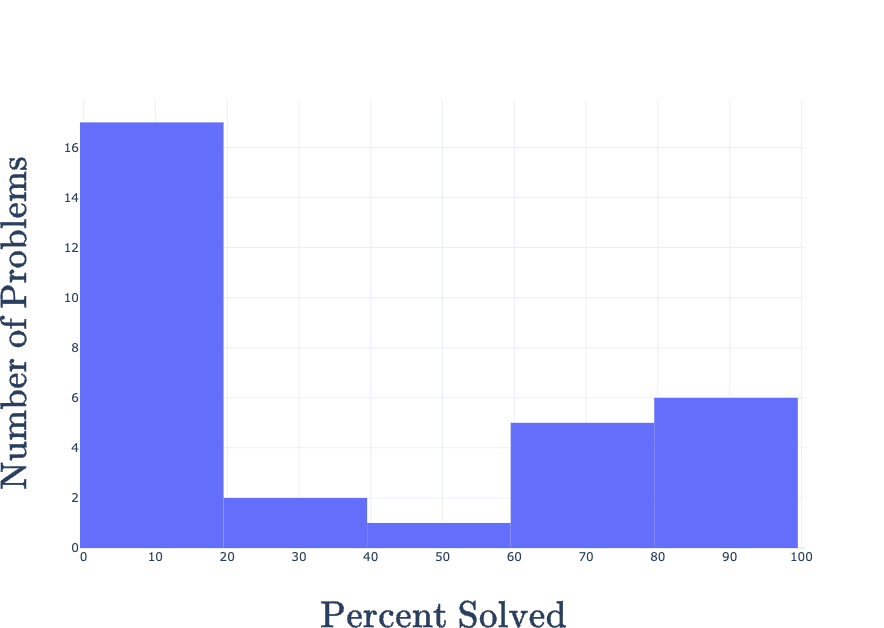}
        \caption{Easy Problems}
        \label{fig:easy_queries}
    \end{subfigure}%
    \hfill
    \begin{subfigure}{0.32\linewidth}
        \centering
        \includegraphics[width=\linewidth]{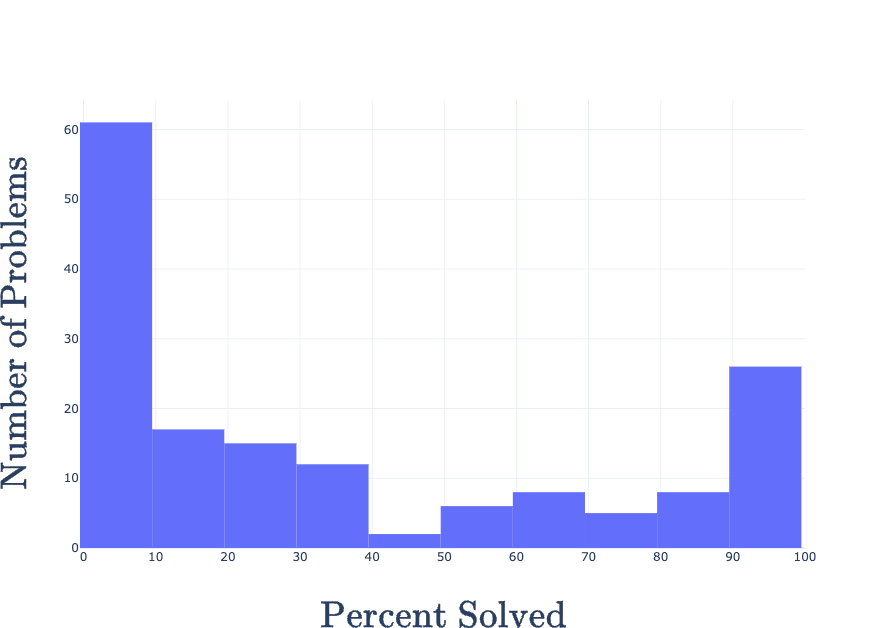}
        \caption{Medium Problems}
        \label{fig:medium_queries}
    \end{subfigure}%
    \hfill
    \begin{subfigure}{0.32\linewidth}
        \centering
        \includegraphics[width=\linewidth]{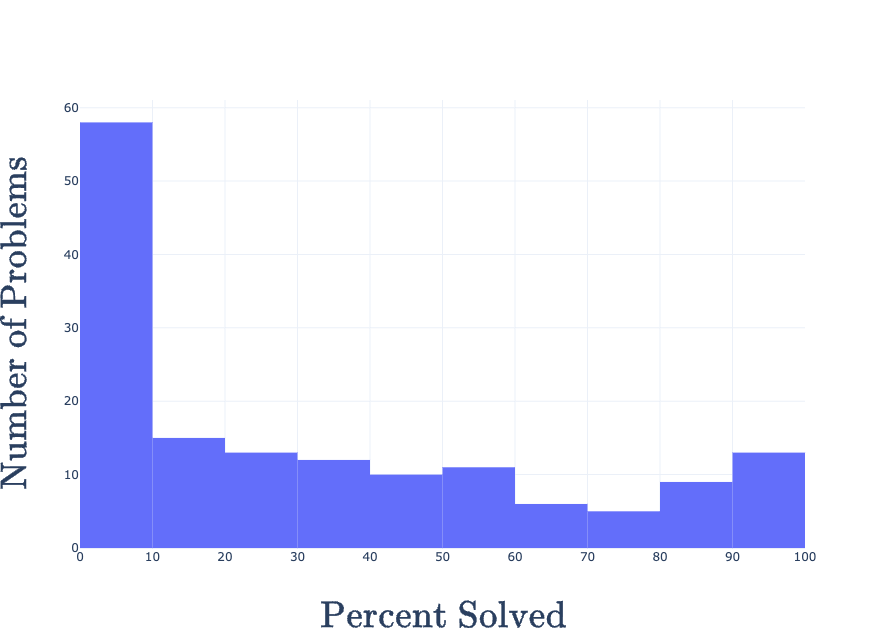}
        \caption{Hard Problems}
        \label{fig:hard_queries}
    \end{subfigure}
    \caption{Distribution of problems not solved by ChatGPT}
    \label{fig:distribution_queries}
\end{figure}

Below is a distribution chart showing GPT-3.5-turbo’s performance across the different difficulty levels, highlighting the problems it could not fully solve. As the difficulty increased, the distribution of unsolved problems became increasingly right-skewed, confirming that GPT-3.5-turbo's ability to solve all test cases diminished with harder problems. For problems where GPT-3.5-turbo was close to solving all test cases, a higher percentage of test cases passed. Notably, there was a significant drop in the number of problems GPT-3.5-turbo could not solve between the easy and the medium/hard categories. While GPT-3.5-turbo struggled similarly with unsolved medium and hard problems (with a failure rate between 0\% and 10\%), it was closer to fully solving more medium problems (90\%-99\%) compared to hard problems.
Figure~\ref{fig:distribution_queries} shows these findings.

\subsection{Hypothesis 2 -- Prompt Engineering:}

We tested three methods to improve GPT-3.5-turbo problem solving performance: chain-of-thought prompt engineering, incorporating failed test cases, and switching to GPT-4. Table~\ref{tab:three_by_four} summarizes the improvements observed across different difficulty levels (Easy, Medium, Hard).

Chain-of-thought (CoT) prompt engineering proved especially effective for easy problems, where it resulted in a 29\% improvement—more than any other method in this category. This aligns well with our hypothesis that simpler problems benefit the most from better-structured prompts. By guiding the model to first generate pseudocode, CoT helps break down easier tasks into manageable steps, allowing GPT-3.5-turbo to catch and correct logical errors early on. However, as problem difficulty increased, the effectiveness of CoT diminished. For medium and hard problems, the improvement rates were lower at 19\% and 14\%, respectively. This suggests that while CoT provides a logical framework for tackling straightforward problems, it struggles to handle the complex problem-solving and deeper reasoning required in higher-difficulty cases.

Incorporating failed test cases into the querys showed a notable performance boost for easy problems, with an improvement of 38\%. This increase, while less pronounced than in other categories, still highlights the model's ability to learn from its mistakes even in simpler scenarios. The effect was even more pronounced in medium-difficulty problems, where this method led to an impressive improvement 60\%. By directly addressing incorrect solutions, the model was able to refine its problem solving approach effectively. The balanced complexity of medium difficulty may facilitate learning from errors while significantly benefiting from the additional information provided. For hard problems, the incorporation of failed test cases resulted in a solid improvement 45\%, reinforcing the value of explicitly correcting errors. Interestingly, many of these failed cases were already embedded in the initial prompts, but reintroducing them as key focal points prompted more sophisticated problem-solving pathways. Overall, the best results were observed in medium-difficulty problems, showcasing the most significant gains in performance through the integration of failed test cases. Interestingly, even though many of these failed cases were already in the initial prompts, reintroducing them as key focal points appeared to achieve better results.

Using GPT-4 instead of GPT-3.5-turbo provided significant improvements across all difficulty levels, particularly for harder problems. For hard problems, GPT-4 showed a 52\% improvement, demonstrating its superior ability to handle more complex scenarios. The model upgrades in GPT-4 seem to have directly enhanced problem-solving capacity across the board, with a balanced 33\% improvement in easy problems and a substantial 58\% gain in medium problems.

Unlike CoT prompt engineering or failed test cases, which are highly dependent on prompt structure and feedback, the improvements brought by switching to GPT-4 highlight the impact of model architecture. While CoT and incorporating failed test cases addressed specific reasoning flaws, GPT-4 appears to offer a more robust, general-purpose upgrade, especially beneficial in the hardest problem category where logical depth and model capacity matter most. In conclusion, we found that prompt engineering with failed test cases yielded results comparable to those of GPT-4. However, the key advantage of GPT-4 is that it does not require additional refining in the queries.

After observing the initial improvements from prompt engineering with GPT-3.5-turbo and from GPT-4, we expanded the scope of our study to include other large language models: Claude 3 Sonnet and Gemini 1.0 Pro. Using the same subset of 20 unsolved and 10 solved problems, we tested these models with the same queries. The results are shown in Table~\ref{tab:three_by_four}.

Claude 3 Sonnet showed strong performance, particularly in the Easy and Medium categories, with improvements of 36\% and 48\%, respectively. However, its performance dropped off in the Hard category, where it improved only by 26\%. This suggests that Claude 3 Sonnet is well-suited to problems of moderate complexity but may struggle with more challenging problems compared to GPT-4, which showed better performance across the board.

Gemini 1.0 Pro, on the other hand, displayed limited improvement, particularly in harder problems. While it showed respectable gains in the Easy category (34\%), its performance in Medium and Hard problems was significantly lower, with only a 17\% and 6\% improvement, respectively. This highlights Gemini 1.0 Pro’s limitations in handling more complex problem-solving tasks.

Overall, the results indicate that while alternative models like Claude 3 Sonnet and Gemini 1.0 Pro can enhance performance, GPT-4 remains the most consistent across all problem difficulty levels. The gap between models becomes more apparent as the problem difficulty increases, with GPT-4 demonstrating superior handling of the more complex tasks.

Overall, the findings support our hypothesis that combining diverse test cases, effective prompt engineering, and model advancements can significantly enhance ChatGPT's performance across a range of problem difficulties as shown in Table~\ref{tab:three_by_four}.

\begin{table}[H]
  \centering
  \begin{tabular}{|c|c|c|c|}
    \hline
     & Easy & Medium & Hard \\
    \hline
    Prompt Engineering & 29\% & 19\% & 14\% \\
    \hline
    Providing Failed Test Cases & 38\% & 60\% & 45\% \\
    \hline
    GPT-4 & 33\% & 58\% & 52\% \\
    \hline
    Claude 3 Sonnet & 36\% & 48\% & 26\% \\
    \hline
    Gemini 1.0 Pro & 34\% & 17\% & 6\% \\
    \hline
  \end{tabular}
  \caption{Evaluating Alternative Prompt Approaches and More Capable LLMs (Percentages illustrate the increase in the number of test cases compared to the baseline)}
  \label{tab:three_by_four}
\end{table}

\subsection{Hypothesis 3 -- Other Languages:}

Lastly, we tested GPT-3.5-turbo across multiple programming languages, specifically C++, Java, Elixir, Erlang, and Racket. For this analysis, we selected a subset of 20 problems: 10 that GPT-3.5-turbo successfully solved in Python and 10 that it could not solve. The results revealed distinct patterns in performance across these languages.

In Elixir, Erlang, and Racket, none of the 20 problems were solved by GPT-3.5-turbo. This highlights potential limitations in the model’s ability to handle these languages effectively. It suggests that the model may struggle with the unique syntax and paradigms of functional programming or may lack sufficient training data in these languages to generate valid solutions.

\begin{figure}[H]
    \centering
    \begin{subfigure}{0.5\linewidth}
        \centering
        \includegraphics[width=\linewidth]{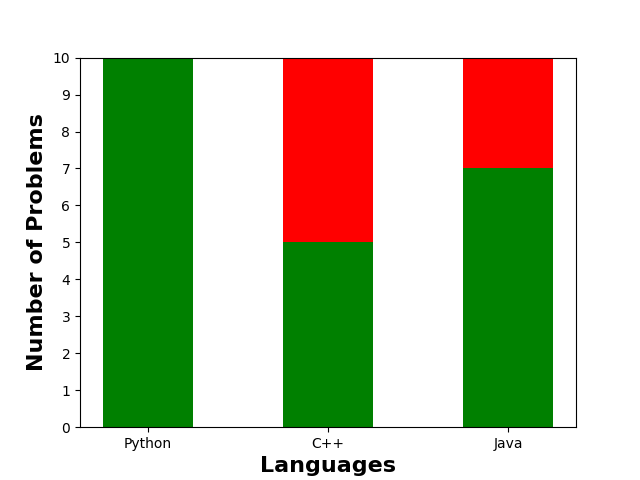}
        \caption{Problems Solved in Python}
        \label{fig:plot_accepted}
    \end{subfigure}%
    \begin{subfigure}{0.5\linewidth}
        \centering
        \includegraphics[width=\linewidth]{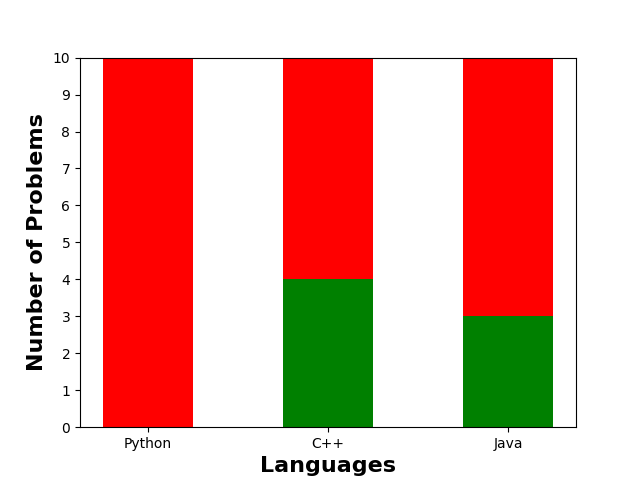}
        \caption{Problems Not Solved in Python}
        \label{fig:plot}
    \end{subfigure}
    \caption{Other Languages Results}
    \label{fig:overall_plots}
\end{figure}

Conversely, C++ and Java produced more meaningful results. We initially assumed that these two languages would exhibit similar performance levels, given the success of Python solutions. However, as figure~\ref{fig:overall_plots} shows C++ solved only 5 out of the 10 problems that ChatGPT successfully solved in Python, while Java solved 7 out of 10 (Note that we omit Elixir, Erlang and Racket from the figure since they could not solve any of the problems). This discrepancy suggests that language-specific challenges play a role in the model’s effectiveness.

Interestingly, in the 10 problems that ChatGPT (GPT-3.5-turbo) could not solve in Python, both C++ and Java managed to solve a few. In C++, the model solved 4 of these 10 problems, with all solutions related to array-based problems. This may indicate that the model's understanding of arrays is better represented in C++ than other topics. However, the more complex syntax and manual memory management of C++ could have limited the model's success, making it less effective overall compared to Python. In Java, GPT-3.5-turbo successfully solved 3 of the 10 previously unsolved problems. These included both arrays and dynamic programming topics. Java’s stronger performance may be due to its explicit handling of data structures and clear object-oriented principles, which align well with the model's training. Furthermore, Java’s verbose syntax likely provides clearer contextual cues, allowing the model to generate more accurate solutions.

The differences in performance between C++ and Java highlight the varying challenges that each language presents. C++, with its focus on pointers and memory management, can hinder the model’s ability to translate logical reasoning into code. In contrast, Java's structured environment and well-defined syntax might make it easier for the model to solve more complex problems, such as those involving dynamic programming. Overall, these results suggest that the effectiveness of large language models like GPT-3.5-turbo varies significantly across programming languages. This variability emphasizes the importance of further research into language-specific training and refining model performance to better accommodate the nuances of each language.

\subsection{Further Results 1 -- Types of Problems:}
After analyzing our initial three hypotheses, we decided to conduct some more experiments. We first decided to look at the types of problems GPT-3.5-turbo did best in. We categorized the problems by type to assess how GPT-3.5-turbo's performance varied across algorithms. As shown in the table below, GPT-3.5-turbo performed best on hash table, search, and divide-and-conquer problems, with only 15-20\% of problems unsolved. These categories typically involve straightforward, modular patterns, making them easier for the model to recognize and generate correct solutions.

However, GPT-3.5-turbo struggled more with database, dynamic programming (DP), and greedy algorithms, leaving 27-31\% of these problems unsolved. These problem types present additional challenges. Database problems require correct SQL syntax and logic, where small errors can invalidate solutions; dynamic programming problems demand state management and optimal substructure reasoning, which can overwhelm the model; and Greedy algorithms often rely on problem-specific heuristics, which the model may struggle to infer correctly.

GPT-3.5-turbo performs well on problems with clear, reusable structures, like search and divide-and-conquer. However, it struggles with complex reasoning and syntax-heavy tasks such as SQL queries and dynamic programming. This variability highlights opportunities for future improvements, particularly in areas that require advanced logic and precision. Table~\ref{tab:mytable} shows our findings.

\begin{table}[H]
    \centering
    \resizebox{\textwidth}{!}{%
    \begin{tabular}{|c|c|c|c|c|c|c|}
        \hline
        \textbf{Algorithms:} & Hash Table & Database & Dynamic Programming & Greedy & Search & Divide and Conquer \\
        \hline
        \textbf{Percent not solved:} & 18\% & 31\% & 30\% & 27\% & 19\% & 15\% \\
        \hline
    \end{tabular}%
    }
    \caption{Types of Problems ChatGPT Couldn't Solve}
    \label{tab:mytable}
\end{table}

\subsection{Further Results 2 -- Lines of Code:}

After seeing the results for hypothesis 1, we further hypothesized that the number of lines of code of in a program correlated to the likelihood the program would be correct. The less lines of code there are, the more likely a program is correct and fulfills the task of the problem. 

\begin{figure}[H]
    \centering
    \begin{subfigure}{0.32\textwidth}
        \centering
        \includegraphics[width=\linewidth]{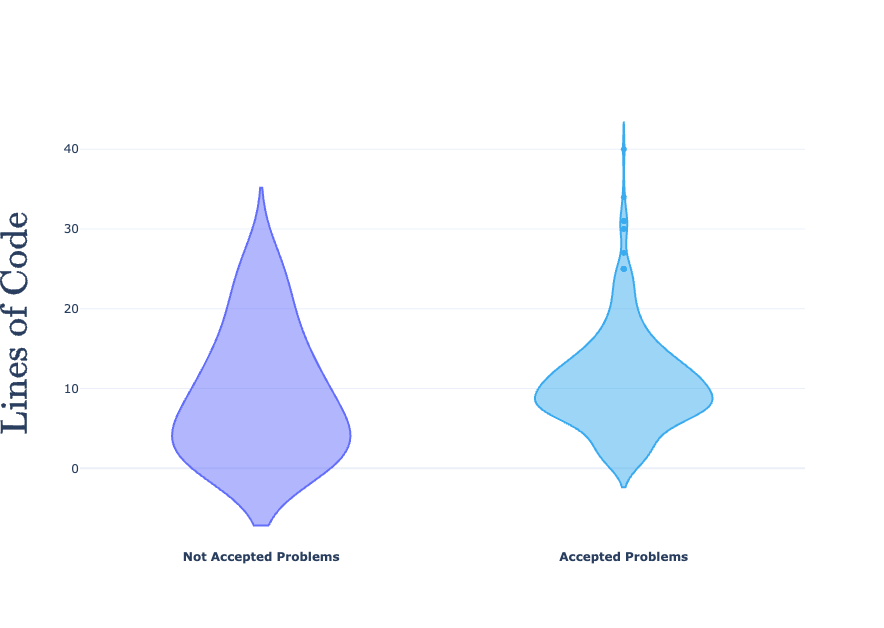}
        \caption{Easy Problems}
        \label{fig:fig1}
    \end{subfigure}%
    \begin{subfigure}{0.32\textwidth}
        \centering
        \includegraphics[width=\linewidth]{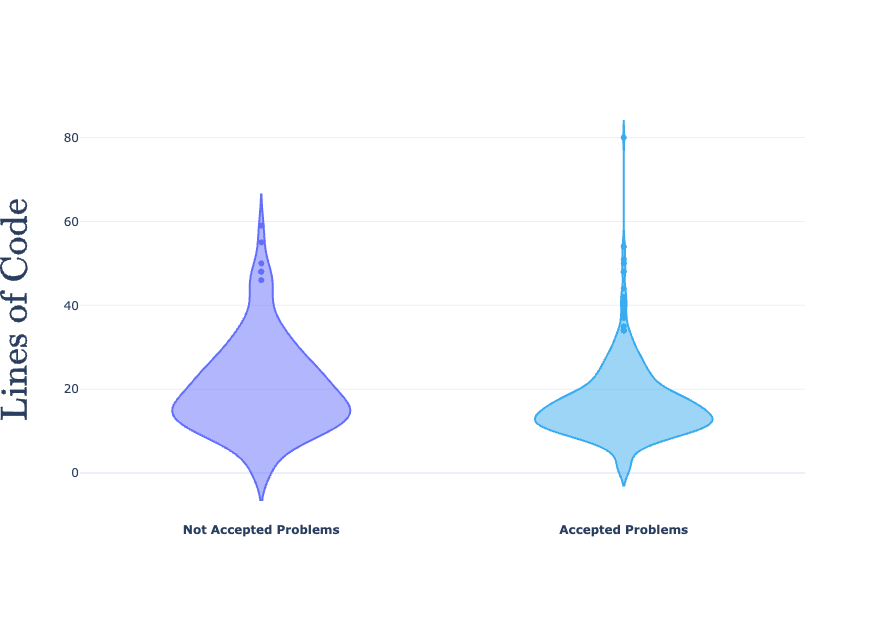}
        \caption{Medium Problems}
        \label{fig:fig2}
    \end{subfigure}%
    \begin{subfigure}{0.32\textwidth}
        \centering
        \includegraphics[width=\linewidth]{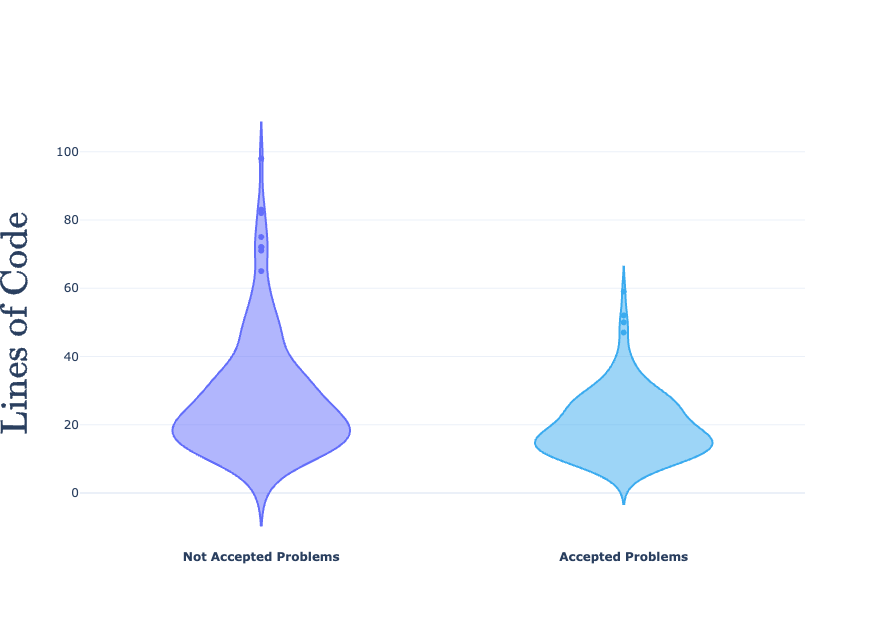}
        \caption{Hard Problems}
        \label{fig:fig3}
    \end{subfigure}
    
    \caption{Lines of Code in Each Difficultly Level}
    \label{fig:overall}
\end{figure}

Our findings support this hypothesis shown in figure~\ref{fig:overall}, although with some nuanced observations. In easy problems, the average lines of code for both solved and unsolved cases were under 10. However, the small difference between the two groups can largely be attributed to MySQL-related tasks, which were shorter in code but frequently unsolvable by GPT-3.5-turbo. These results suggest that for simpler tasks, efficiency in code length was not the primary factor driving correctness.

At the medium difficulty level, the hypothesis appeared to hold more strongly: Solved problems averaged around 15 lines of code, while unsolved ones average around 17 lines of code. Surprisingly, there were outliers where GPT-3.5-turbo generated correct solutions exceeding 50 lines of code, indicating that the model sometimes relied on brute force approaches instead of efficient algorithms. This suggests that while shorter code generally led to correct solutions, the model occasionally resorted to longer, less efficient methods to solve more complex scenarios.

The most notable trend emerged in hard problems. Solved problems averaged only 15 lines, while unsolved problems had an average of 20 lines of code. Additionally, there were many unsolved problems with more than 50 lines of code, highlighting GPT-3.5-turbo’s difficulty with excessively long or complex programs. These results suggest that the model’s strength lies in maintaining concise and efficient code even as the problem difficulty increases, and that it tends to perform worse when solutions become too lengthy.

These findings reinforce the idea that concise code is generally more effective, especially for harder problems. While GPT-3.5-turbo performed well with shorter solutions, outliers in medium and hard problems show that the model sometimes relied on brute force methods. The results emphasize the importance of efficiency in the length of the code, as excessive code appears to diminish the success rate of the model, particularly for more difficult problems.

\FloatBarrier 


\section{Conclusions \label{sec:conclusions}}

This research investigated ChatGPT’s (GPT-3.5-turbo) performance in solving coding problems of varying complexity, across different programming languages, and with the application of prompt engineering techniques. Our findings provide valuable insights into the capabilities and limitations of ChatGPT in automated code generation. Table~\ref{tab:overall-findings} shows a summary of our findings. 

\begin{table}[ht]
\centering
\begin{tabularx}{\textwidth}{|>{\raggedright\arraybackslash}X|>{\raggedright\arraybackslash}X|}
    \hline
    \textbf{Research Question} & \textbf{Results} \\ \hline
    How effectively can ChatGPT solve Leetcode problems across varying levels of difficulty (easy, medium, and hard)? 
    & ChatGPT’s performance declines from easy to hard problems, performing best on easy and worst on hard. \\ \hline 
    Can prompt engineering and the selection of more effective demonstrations enhance ChatGPT's initial coding results? 
    & Prompt engineering showed better results by including test cases ChatGPT initially failed in the query. \\ \hline
    Which programming languages is ChatGPT most proficient in? 
    & ChatGPT demonstrates stronger performance in Python and Java, with more difficulty in compiled languages like C++ and newer languages. \\ \hline
\end{tabularx}
\caption{Summary of overall findings related to ChatGPT's coding performance}
\label{tab:overall-findings}
\end{table}

We observed a clear correlation between the difficulty of the problem and ChatGPT's success rate. As the complexity of the problems increased, the model's ability to solve them decreased significantly. This outcome was consistent with our hypothesis that ChatGPT’s performance would diminish on medium and hard problems. Additionally, our analysis of the number of lines of code revealed that shorter, more efficient solutions were more likely to be correct, though brute-force solutions emerged in some cases, particularly for medium and hard problems.

Furthermore, our exploration of prompt engineering techniques demonstrated the potential to improve performance, with putting failed test cases into the query yielded the most substantial gains. Although GPT-4 offered enhanced capabilities, particularly in the medium and hard problem categories, it was clear that thoughtful prompting, combined with the introduction of diverse test cases, contributed meaningfully to problem solving improvements. These findings highlight the importance of input structuring when working with large language models in coding applications.

Our comparative analysis of ChatGPT performance in different programming languages revealed interesting disparities. While Python remained the strongest language, models faced greater challenges in languages like C++ and Java. Moreover, programming languages such as Elixir, Erlang, and Racket posed considerable obstacles, with no solutions generated, emphasizing the need for further exploration of language-specific optimizations for these models. Lastly, analysis of the types of problems identified the strengths of ChatGPT in solving hash table, search, and divide-and-conquer problems, while dynamic programming and database-related problems proved to be more challenging.

In conclusion, while ChatGPT demonstrates impressive problem-solving capabilities, particularly for simpler problems, there are significant limitations in handling complex algorithms, certain languages, and problem types. Our study emphasizes the role of tailored prompt engineering in enhancing the performance of these models. As large language models continue to evolve, future research should focus on overcoming these challenges and refining the behavior of the model to improve the reliability of the coding in more advanced scenarios.



\bibliographystyle{abbrv}
\bibliography{references} 

\end{document}